\documentclass{elsart5}

 \usepackage{graphicx}

\usepackage{amssymb}

\usepackage{amsmath}
\usepackage{graphicx,epsfig}
\usepackage{subfigure}

\newcommand{\be}{\begin{equation}}
\newcommand{\bd}{\begin{displaymath}}
\newcommand{\ed}{\end{displaymath}}
\newcommand{\ee}{\end{equation}}
\newcommand{\nn}{\nonumber\\}
\newcommand{\ij}{\langle ij \rangle}

\newcommand{\ba}{\begin{eqnarray}}
\newcommand{\ea}{\end{eqnarray}}

\begin{document}

\begin{frontmatter}

\title{Classical Ring-exchange Processes on the Triangular Lattice
\thanksref{tit1}}
\thanks[tit1]{This work is supported by Korea Research
Foundation through Grant No. KRF-2005-070-C00044.}

\author[aff1]{June Seo Kim}
\corauth[cor1]{}
\author[aff1,aff2]{Jung Hoon Han \corauthref{cor1}}
\ead{hanjh@skku.edu} \ead[url]{http://manybody.skku.edu}
\address[aff1]{Department of Physics and Institute for
Basic Science Research, \\
Sungkyunkwan University, Suwon 440-746, Korea}
\address[aff2]{CSCMR,
Seoul National University, Seoul 151-747, Korea}


\begin{abstract}
The effects of the ring-exchange Hamiltonian $H_3 =  {J_3 }
\sum_{\langle ijk \rangle} (S_i \cdot S_j) (S_i \cdot S_k )$ on the
triangular lattice are studied using classical Monte Carlo
simulations. Each spin $S_i$ is treated as a classical XY spin
taking on $Q$ equally spaced angles ($Q$-states clock model). For
$Q=6$, a first-order transition into a stripe-ordered phase preempts
the macroscopic classical degeneracy. For $Q > 6$, a finite window
of critical phase exists, intervening between the low-temperature
stripe phase and the high-temperature paramagnetic phase.
\end{abstract}

\end{frontmatter}


The importance of higher-order ring-exchange processes in
low-dimensional magnets and its potential role in stabilizing
liquid-like exotic ground states has recently been under intense
investigation\cite{lhuillier,motrunich}. Generically the Hamiltonian
is of the type

\be  H = \sum_{n\ge 2} H_n \label{Hn}\ee
with $H_n$ for $n=2,3,4$ given by
\ba && H_2 = J_2 \sum_{\ij} S_i \cdot S_j \nn
&& H_3 =  {J_3 } \sum_{\langle ijk \rangle} (S_i \cdot S_j) (S_i
\cdot S_k ) \nn
&& H_4 = J_4 \sum_{\langle ijkl \rangle} (S_i \cdot S_j ) (S_k \cdot
S_l ).
\label{Hamiltonian}\ea
Each $S_i$ is a Heisenberg spin, and $\ij$, $\langle ijk \rangle$
and $\langle ijkl \rangle$ refer to nearest-neighbor pair, triplet,
and quartet of sites, respectively.

In particular the possibility to realize a stable spin-liquid ground
state due to $H_3$ in a two-dimensional triangular lattice has been
suggested in the variational Monte Carlo study of
Motrunich\cite{motrunich}. It may be inferred that the three-site
exchange process has the ``frustrating" effect which renders the
liquid ground state  energetically more stable over a
$\sqrt{3}\times\sqrt{3}$ magnetically ordered structure.

In this paper, we report the first results of isolating the effects
of the three-site exchange process by studying the following model.
First, we consider the classical counterpart of the Hamiltonian
(\ref{Hn}) where $S_i$ is treated as a uni-modular vector. Secondly,
only $H=H_3$ is considered in this paper while leaving the study of
the compound models such as $H=H_2 + H_3 $ or $H = H_2 + H_3 + H_4$
for the future. Thirdly, we consider the planar spin $S_i = (\cos
\theta_i, \sin\theta_i )$. The angle $\theta_i$ is divided up into
$Q$ equally spaced segments. The same strategy had been applied for
$H=H_2$ as a way to asymptotically approach the behavior of the XY
model ($Q=\infty$) and is known as the $Q$-states clock
model\cite{grest,landau}.

Antiferromagnetic $Q$-states models on a triangular lattice have
double transitions of XY- and Ising-types with extremely close
critical temperatures\cite{noh}. The same subtlety might pervade the
$H=H_3$ model too, but here we choose to focus on the broader issue:
\textit{ What is the nature of the low-temperature phase exhibited
by $H=H_3$? }  The results for $Q=2$ and $Q=6$ are discussed in
detail in this paper. Some preliminary specific heat data for larger
$Q$ are presented.
\\

\textbf{Q=2}: It turns out that $Q=2$ model maps onto the
antiferromagnetic Ising model on the triangular net, first studied
by Wannier\cite{wannier}.  In the Ising case spins take on $S_i =
\pm 1$, and the Hamiltonian $H_3$ reduces to

\be   (S_i \cdot S_j ) (S_i \cdot S_k ) \rightarrow S_j S_k, ~~~ H_3
\rightarrow 2 J_3 \sum_{\ij} S_i S_j ,\ee
which is the antiferromagnetic Ising model. This model possesses
macroscopic degeneracy\cite{wannier,anderson} which is also revealed
as the residual entropy $S_0$.  Our Monte Carlo (MC) calculation
gives $S_0 \approx 0.323 k_B$, in excellent agreement with the value
predicted earlier\cite{wannier,anderson}. There is no long-range
order down to zero temperature in this model.

With $Q \geq 3$ $H_2$ and $H_3$ are no longer equivalent. The
lowest-energy configurations for $H_3$ consist of two-up and
one-down (or vice versa) spins for each elementary triangle. Thus,
macroscopic degeneracy is a general feature of $H=H_3$ for an
arbitrary even integer $Q$. It is also well known that the magnetic
ordering for $H=H_2$ is obtained for  angles of 120$^\circ$ between
nearest-neighbor spins. Such situations are possible  if $Q$ is a
multiple of 3. To allow the realization of both, we consider the
case $Q=6$.
\\

\begin{center}
\begin{figure}[h]
\includegraphics[angle=0,width=9.0cm]{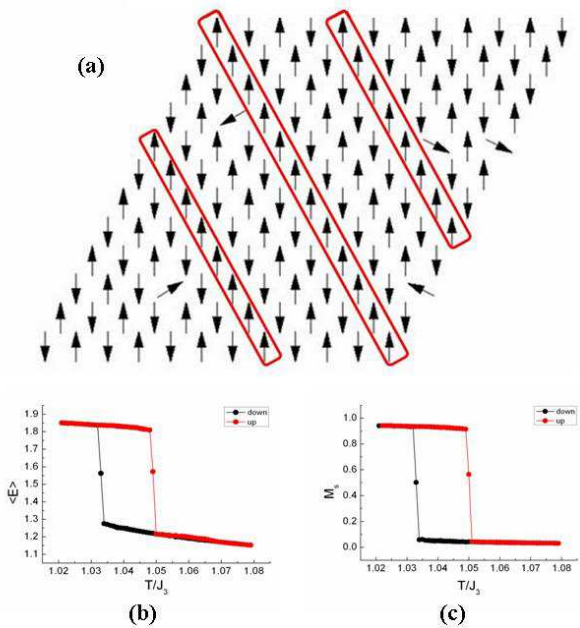}
\caption{Low-temperature configuration for the $Q=6$ model, $H=H_3$.
Ferromagnetic ordering within a diagonal stripe and
antiferromagnetic ordering between adjacent stripes are apparent in
(a). Both the average energy (b) and the average magnetization (c)
data are consistent with the first-order phase transition to the
low-temperature phase.} \label{spontaneous-stripe}
\end{figure}
\end{center}

\textbf{Q=6}: With $Q=6$ we observed a \textit{first-order}
transition at $T_c /J_3 \approx 1.05$. Hysteresis in the average
energy and the order parameter (defined below) in our MC runs
vindicate the first-order nature, as shown in Fig.
\ref{spontaneous-stripe}.

The nature of the low-temperature, ordered phase is clearly
demonstrated in Fig. \ref{spontaneous-stripe}. We find the
spontaneous emergence of stripe-like domains of up and down spins
below $T_c$. The ground state degeneracy is lifted through the
order-from-disorder mechanism\cite{villain,henley}. While a
conventional order-from-disorder idea predicts the gradual
separation of the free energies of different classical ground state
configurations with rising temperatures and no phase transition, our
model exhibits a first-order transition which pre-empts the
classical macroscopic degeneracy. The difference is due to the
discrete nature of our model. The order parameter appropriate for
this stripe-like configuration is

\be m = {1\over N} \left| \sum_{i} (-1)^{i_1} S_i  \right|, \ee
where each lattice site is given the coordinate $i = i_1 \hat{e}_1 +
i_2 \hat{e}_2$, $\hat{e}_1 = \hat{x}$, $\hat{e}_2 = -\hat{x}/2 +
\sqrt{3}\hat{y}/2$, and $N$ is the number of sites.
\\

\textbf{Q$>$6}: For a finer spin segmentation we still obtain the
low-temperature stripe-like phase. A single first-order phase
transition observed for $Q=6$ is split into two transitions, at
temperatures $T_1$ and $T_2$ with $T < T_1$ being the stripe-ordered
phase. The intermediate phase $T_1 \le T \le T_2$ appears to be
critical\cite{landau,in-progress}.

\begin{center}
\begin{figure}[h]
\includegraphics[angle=0,width=8.0cm]{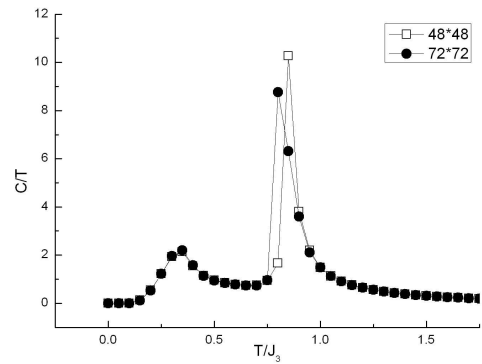}
\caption{Specific heat $C(T)/T$ for $Q=12$, $H=H_3$, for
$48\times48$ and $72\times72$ lattices. The two peaks are indicative
of the presence of two phase transitions. The low temperature phase
is given by the stripe configuration shown in Fig.
\ref{spontaneous-stripe}. } \label{Q12}
\end{figure}
\end{center}

\end{document}